# PublicnEUro: A FAIR and TRUSTworthy infrastructure for controlled access to human neuroimaging data

Cyril R Pernet[1,2], Stephan Heunis[3], Michael Hanke[3,4], Alyssa Dai[5], Jean Baptiste Poline[5,6,7], Peter S Jensen[1], Martin Nørgaard[8], Melanie Ganz[1,8] & Gitte M Knudsen[1,9]

1 Neurobiology Research Unit, Rigshospitalet, Copenhagen, Denmark
2 Department of Health Technology, Technical University of Denmark, Lyngby, Denmark
3 Institute of Neurosciences and Medicine, Jülich, Germany
4 Heinrich Heine University Düsseldorf, Düsseldorf, Germany
5 McConnell Brain Imaging Centre, The Neuro (Montreal Neurological Institute-Hospital), Montreal, Canada
6 Department of Neurology and Neurosurgery, McGill University, Montreal, Canada
7 School of Computer Science, McGill University, Montreal, Canada
8 Department of Computer Science, University of Copenhagen, Copenhagen, Denmark
9 Department of Clinical Medicine, University of Copenhagen, Copenhagen, Denmark

**ORCID**
Cyril R Pernet https://orcid.org/0000-0003-4010-4632
Stephan Heunis https://orcid.org/0000-0003-3503-9872
Michael Hanke https://orcid.org/0000-0001-6398-6370
Alyssa Dai https://orcid.org/0000-0003-0174-6800
Jean Baptiste Poline https://orcid.org/0000-0002-9794-749X
Peter S Jensen https://orcid.org/0000-0001-7058-6930
Melanie Ganz https://orcid.org/0000-0002-9120-8098
Gitte M Knudsen https://orcid.org/0000-0003-1508-6866

## Abstract

Sharing brain imaging data supports secondary analysis and reproducibility, but privacy risks and differing legal and governance requirements can make unrestricted release inappropriate. The PublicnEUro repository was developed to support discovery and reuse under diverse legal requirements. Here, PublicnEUro's governance, architecture, BIDS-based deposition and curation, metadata and identifier services as well as open and controlled retrieval workflows are described. We then assess the repository published datasets against FAIR subprinciples and the repository public information against TRUST principles. The FAIR assessment shows strong findability and public metadata accessibility. The TRUST assessment documents governance, stewardship, continuity provisions and technical safeguards. Together this demonstrates that PublicnEUro, by separating open discovery from governed data retrieval, supports FAIR-aligned and trustworthy reuse of sensitive brain imaging data without requiring unrestricted disclosure.

**Keywords:** PublicnEUro, brain imaging, repository, resource, data sharing, FAIR, TRUST

# 1. Introduction

Sharing data extends the scientific value of an acquisition beyond its original research question. Secondary analyses can combine cohorts, test new hypotheses, develop methods, and independently examine the robustness of published findings (Botvinik-Nezer et al., 2019; Poline et al., 2012). This opportunity is particularly valuable in brain imaging: collecting data requires specialized equipment, participant recruitment, and phenotyping, while modalities such as Positron Emission Tomography (PET) add radiotracer production and acquisition-specific measurements; these costs can constrain sample sizes and make some cohorts difficult to replicate (Knudsen et al., 2020; Milham et al., 2018). Capitalizing on existing datasets therefore requires more than retaining image files. Repositories must preserve discoverable dataset records, persistent identifiers, and sufficient standardized metadata for researchers to understand, cite, and request data for reuse (Gorgolewski et al., 2016; Wilkinson et al., 2016).

For researchers to share their data, repositories need to provide services that address legal, ethical, and intellectual property requirements. Human neuroimaging data are shared within a global landscape of privacy law. Jurisdictions with substantial neuroimaging research activity—including the European Union (EU), United States, Canada, Australia, China, Japan, and South Korea—regulate aspects of identifiable health information, although their definitions and legal requirements differ (DLA Piper Intelligence, 2026). Depending on the jurisdiction, information receiving heightened protection may include race or ethnicity, religion, political beliefs, and health, genetic, or biometric information. For neuroimaging, information about physical and mental health is particularly relevant: brain scans and associated clinical or phenotypic measures may reveal health information, and removing direct identifiers does not necessarily eliminate the possibility of identifying a participant (Kibsgaard et al., 2025; Schwarz et al., 2019). Repositories therefore need to offer appropriate safeguards, including access conditions, records of disclosures, and controls for international transfers where required. Controlled access can also serve purposes beyond legal compliance. First, dataset-specific agreements and approval procedures can apply the permissions and restrictions established for each dataset, helping data authors respect participants' rights and expectations. Second, controlled access can provide additional protection for sensitive or vulnerable populations, including people with rare diagnoses or distinctive attributes (Kibsgaard et al., 2025; Machanavajjhala et al., 2007). Third, an embargo can give authors time to publish planned analyses before a broader data release, while access conditions can protect legitimate intellectual property interests.

In the EU, repositories offer different conditions for depositing and accessing data, reflecting the range of researchers' needs (Lefort-Besnard et al., 2025). PublicnEUro (Heunis & Pernet, 2024) was created as a Research Data Repository (re3data.org, 2026; RRID:SCR_027770) dedicated to brain imaging, offering services for sharing data under varying access requirements, including those arising under the EU General Data Protection Regulation (GDPR). This Technical Report describes PublicnEUro's architecture, governance, and operational workflows, evaluates their alignment with the FAIR (Wilkinson et al., 2016) and TRUST (Lin et al., 2020) principles, and summarizes the results using the open-source PolicyModels toolkit (Murphy et al., 2021).

## 2. Materials and methods

### 2.1. Material

PublicnEUro comprises three interconnected functional layers (Figure 1). (i) The *governance and repository-management layer* defines institutional responsibilities, legal agreements, curation procedures, dataset lifecycle management, and access decisions, and is supported by an operational database. (ii) The *secure data-services layer* provides storage, preservation, authentication, authorization, auditing, and controlled data retrieval within the hosting environment. (iii) The *discovery and interoperability layer* exposes the public catalog, persistent identifiers, dataset-level metadata, machine-readable use conditions, and repository-level representations. Governance spans the two technical layers, while persistent dataset identifiers and status records maintain consistency between protected data holdings and their public representations.

Governance and management: PublicnEUro is operated by the Neurobiology Research Unit (NRU) at Copenhagen University Hospital Rigshospitalet under the governance of the Capital Region of Denmark. It was established as a research infrastructure as one of the aims of the OpenNeuroPET project (Ganz et al., 2026) based on funding by the Novo Nordisk Foundation. For deposited datasets, the originating institution remains the data controller and retains ownership and decision-making authority over permitted uses. NRU acts as data processor by receiving, curating, storing and distributing data according to the controller's documented instructions and the applicable service and data-processing agreements. These services are offered against fees at cost values (recovering admin time and compute/storage costs). PublicnEUro separately acts as controller for the personal information required to administer repository accounts and access records. Governance is divided between an Oversight Committee and a Management Committee. The Oversight Committee provides strategic direction, reviews policies and scientific choices, and monitors repository activity. The Management Committee translates that direction into technical specifications and operating procedures, reviews access applications where oversight is required, audits dataset use and reports repository activity to the Oversight Committee. A Data Protection Officer (DPO) advises both committees, contributes to the Data Protection Impact Assessment (DPIA) and risk register, reviews access records and coordinates regulatory or controller notification when an incident involves personal data.

Dataset-specific sharing rules are implemented in an SQL-backed operational database that connects users, institutions, agreements, datasets, access requests, and audit records. At a conceptual level, three binary policy decisions—whether data access is open or controlled, whether geographical eligibility is restricted, and whether approval is automatic or requires oversight—yield five data-sharing configurations. The implementation provides a finer representation: dataset records that distinguish among private, EU-only, EU-and-adequate-country, worldwide, and open accessibility, as well as automatic or oversight-based approval. Access checks verify user and institutional status, geographical eligibility, approval requirements, and any applicable Standard Contractual Clauses before a temporary retrieval link is issued. The PublicnEUro database implementation documents these rules and the associated browser and command-line services, and is distributed openly under an MIT license. Retention and preservation commitments are defined in each dataset's service agreement rather than through a single universal retention period. At the end of an agreement, the controller and NRU may renew active hosting, move files to cold archival storage, return the curated BIDS dataset to the controller, or remove the files while retaining the public metadata and Digital Object Identifier (DOI) landing page with an updated lifecycle status. Dataset-level succession is therefore managed contractually through instructions for return, deletion, or archival.

Data services: Deposition begins with a contract between the data authors or responsible institution and NRU. Authors remain responsible for lawful collection, appropriate pseudonymization, and provision of the dataset-specific license or Data User Agreement (DUA). Files are transferred through a secure route and must be organized according to the Brain Imaging Data Structure (BIDS, Poldrack et al., 2024), including Magnetic resonance imaging (MRI), PET, electroencephalography (EEG), magnetoencephalography (MEG), and related modalities. The BIDS Validator serves as the primary technical acceptance test, although BIDS-like datasets are also accepted (for instance, a dataset implementing a BIDS Extension that has not yet been released). Dataset-level information is supplied through the PublicnEUro metadata record and supplemented during curation. Enhancement can include correction of BIDS organization and metadata, catalog checks, participant summaries, Neurobagel annotation (Urchs et al., 2026), and translation of the authoritative license or Data User Agreement into machine-readable Digital Use Conditions (DUC; Jeanson et al., 2024). PublicnEUro assigns an internal PublicnEUro (PN) identifier, registers a DOI, and creates the public catalog record. An updated data release can be published as a new version with its own DOI while remaining linked to earlier or superseded versions. Dataset files are hosted in the Computerome hosting environment, PublicnEUro's approved sub-processor for storage and computation. Computerome is based in Denmark and is designed to process sensitive research and health data. The hosting and database service components are operating under ISO 27001-certified controls.

Open datasets can be accessed directly through a browser, although registration is required for command-line retrieval. For controlled datasets, applicants create an account, confirm their email address and institutional affiliation, and undergo identity verification by the PublicnEUro team. The access service also evaluates the institution's country against the dataset's geographical policy. Transfers outside the EU/EEA and to countries covered by an adequacy decision require the applicable EU Standard Contractual Clauses to be signed by the user institution before access is enabled. Access is requested separately for each dataset and requires acceptance or signature of its Data User Agreement. Depending on the controller's policy, a compliant request may be approved automatically upon online agreement, or it may require that the signed agreement be reviewed and approved by the controller or an authorized administrator. After approval, the system generates a time-limited browser retrieval link; registered and authorized users can also retrieve a complete dataset or selected folders through the command-line interface. Requests, approvals and subsequent access are recorded so that the data controller can audit who obtained which dataset and under which conditions.

PublicnEUro combines organizational and technical controls to protect confidentiality, integrity, availability and accountability. Individual accounts, manual registration checks, password authentication and one-time links support identity and access management. Authorization is dataset-specific and can be revoked; staff access is limited to operational need and individual actions are logged. Review of authorizations and DPO oversight connect technical controls to GDPR responsibilities. Incident procedures require NRU to notify the relevant data controller without unnecessary delay when confidentiality, integrity, or availability may have been compromised, and to provide the information needed for regulatory assessment, containment, and recovery.

Discovery and interoperability: The public catalog is generated by representing each file and folder of a dataset as JSON files and combining the metadata in a dataset-level file using the DataLad data catalog (Halchenko et al., 2021). This is then automatically rendered as a dataset-specific landing page that displays all the information (which is also publicly available on GitHub). The dataset-level JSON records present descriptive, technical, participant, funding, authorship, license or access condition, lifecycle, and retrieval metadata without exposing participant-level files. Every dataset receives a stable PN identifier, and each published version is represented explicitly with different Crossref DOIs. Version records are marked active, archived, retired, withdrawn or superseded independently. The status is separated from retrieval mode so that continued curation, physical storage, and the current access route remain distinguishable. Users can search and browse the catalog using dataset descriptions, authors, keywords, and structured metadata, while the underlying files remain protected where required, or they can use the Neurobagel query tool searching for imaging data type, disease status, age ranges, etc for participants across datasets, on PublicnEUro or even across repositories linked via Neurobagel.

The PublicnEUro metadata repository derives a governance record for every dataset and version from the catalog. These records preserve the source path and content hash, lifecycle status, retrieval information, DOI, DUC profile, and provenance of the licence-to-DUC mapping. JSON Schema governs the dataset representation, while repository-wide re3data and OpenAIRE CERIF XML exports support exchange with external information systems.

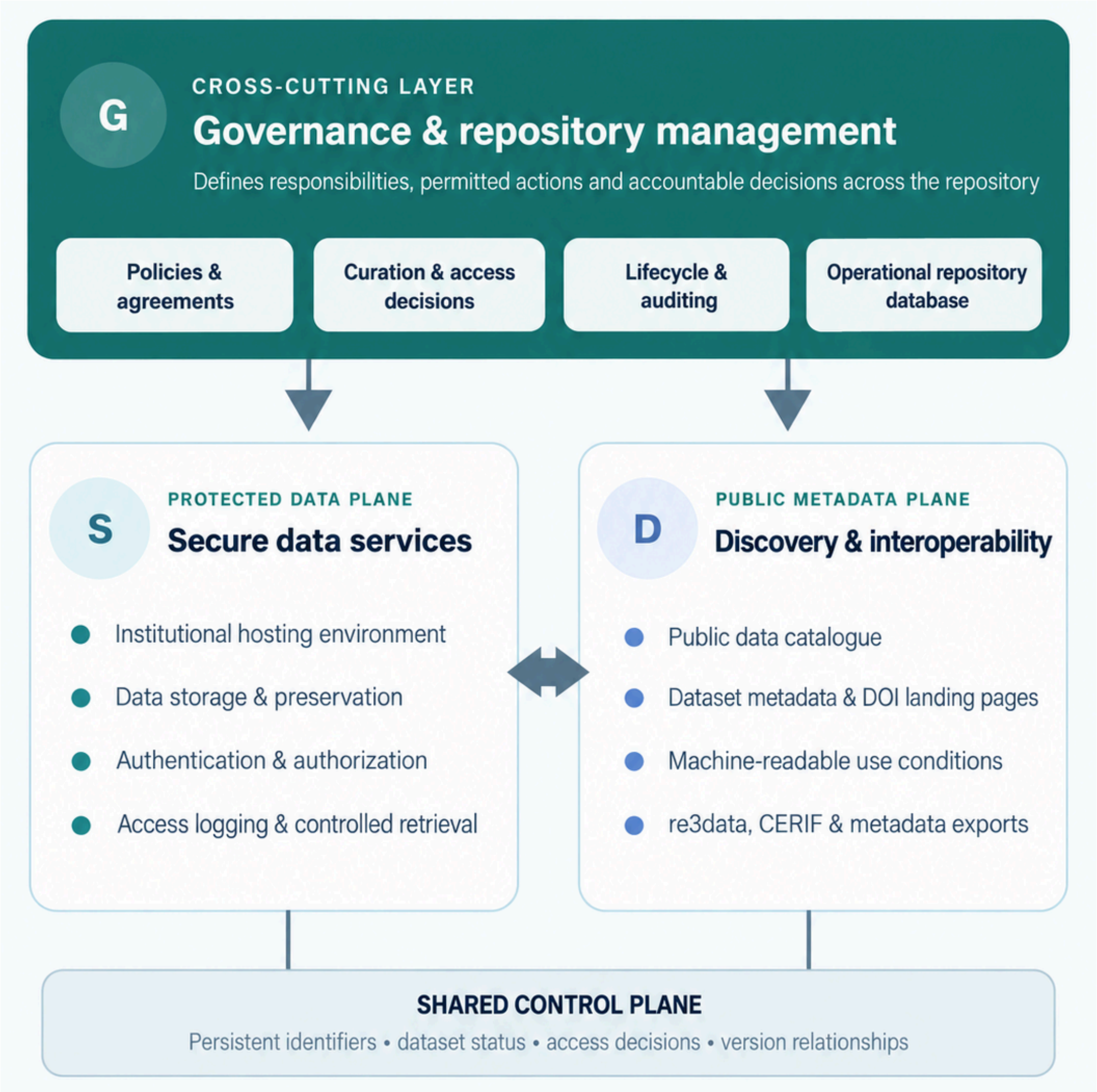

***Figure 1. Functional architecture of PublicnEUro.*** *The governance and repository-management layer defines policies, responsibilities, curation procedures, lifecycle management, and access decisions across the repository. The secure data-services layer provides protected storage, preservation, authentication, authorization, logging, and controlled retrieval. The discovery and interoperability layer exposes the public catalog, persistent identifiers, dataset metadata, machine-readable use conditions, and repository-level metadata exports. Persistent identifiers, dataset status, access decisions, and version relationships maintain consistency between the protected data and public metadata planes.*

## 2.2 Method

**FAIR assessment:** FAIR alignment was evaluated against the 15 subprinciples of the FAIR Guidelines using a structured self-assessment. The unit of assessment was a published dataset version together with its public metadata and access mechanism. Evidence was taken from the PublicnEUro website FAQ, the policy and standards repository (also available on the website), the PublicnEUro DataLad catalog and DOI landing pages, and the dataset-level JSON Schema, JSON records, and repository-wide XML exports. Each subprinciple was rated in the following way:

- Principle demonstrated when the required behavior was both documented and inspectable,
- Partially demonstrated when implementation or evidence was incomplete or internally inconsistent, and
- not demonstrated when no public evidence was identified.
- Ratings concern the findability, accessibility, interoperability, and reusability of data and metadata, not scientific validity or data quality. Authentication and authorization for sensitive data were evaluated under A1.2 and not treated as accessibility failures.

**TRUST assessment:** TRUST alignment was evaluated against the five TRUST principles: Transparency, Responsibility, User focus, Sustainability, and Technology (Lin et al., 2020) using a structured self-assessment. The unit of assessment was PublicnEUro as an organizational and technical repository service rather than individual dataset versions. The requirements described for each principle were operationalized as evidence questions covering public disclosure of mission, scope, terms, and preservation commitments; assignment and performance of stewardship responsibilities; alignment with and responsiveness to the target community; funding, continuity, risk, and succession planning; and use of appropriate standards, infrastructure, and security controls. Evidence was taken from the PublicnEUro website and FAQ, its version-controlled policy and standards repository, the public DataLad catalog, dataset-level governance metadata and repository-level XML exports. Each principle was rated demonstrated when the relevant capability and supporting public evidence were sufficiently complete, partial when evidence or implementation was incomplete, and not demonstrated when no public evidence was identified.

**Open, FAIR, Citable and Trustworthy (OFCT) summary:** Following the principle-based FAIR and TRUST assessments, we answered the 31-question OFCT decision-tree instrument developed by Murphy et al. (2021). The OFCT instrument operationalizes repository characteristics as observable criteria, adds explicit dimensions for openness and data citability, and produces structured profiles that identify documentation and implementation gaps.

# 3. Results

As of October 2026, PublicnEuro is hosting 24 datasets (25 versions), counting 2130 participants for a total size of 5.07 TB. The catalogue records represent MRI, PET, EEG, and CT neuroimaging data. MRI includes structural and functional imaging, perfusion, and diffusion data. There are 31 verified users from institutions across the world and we are currently registering around 5 users a month. Datasets requiring oversight are typically approved in 2 weeks turnaround. Figure 2 displays the geographical distribution of downloaded access controlled datasets, and therefore registered users (open access data are not tracked).

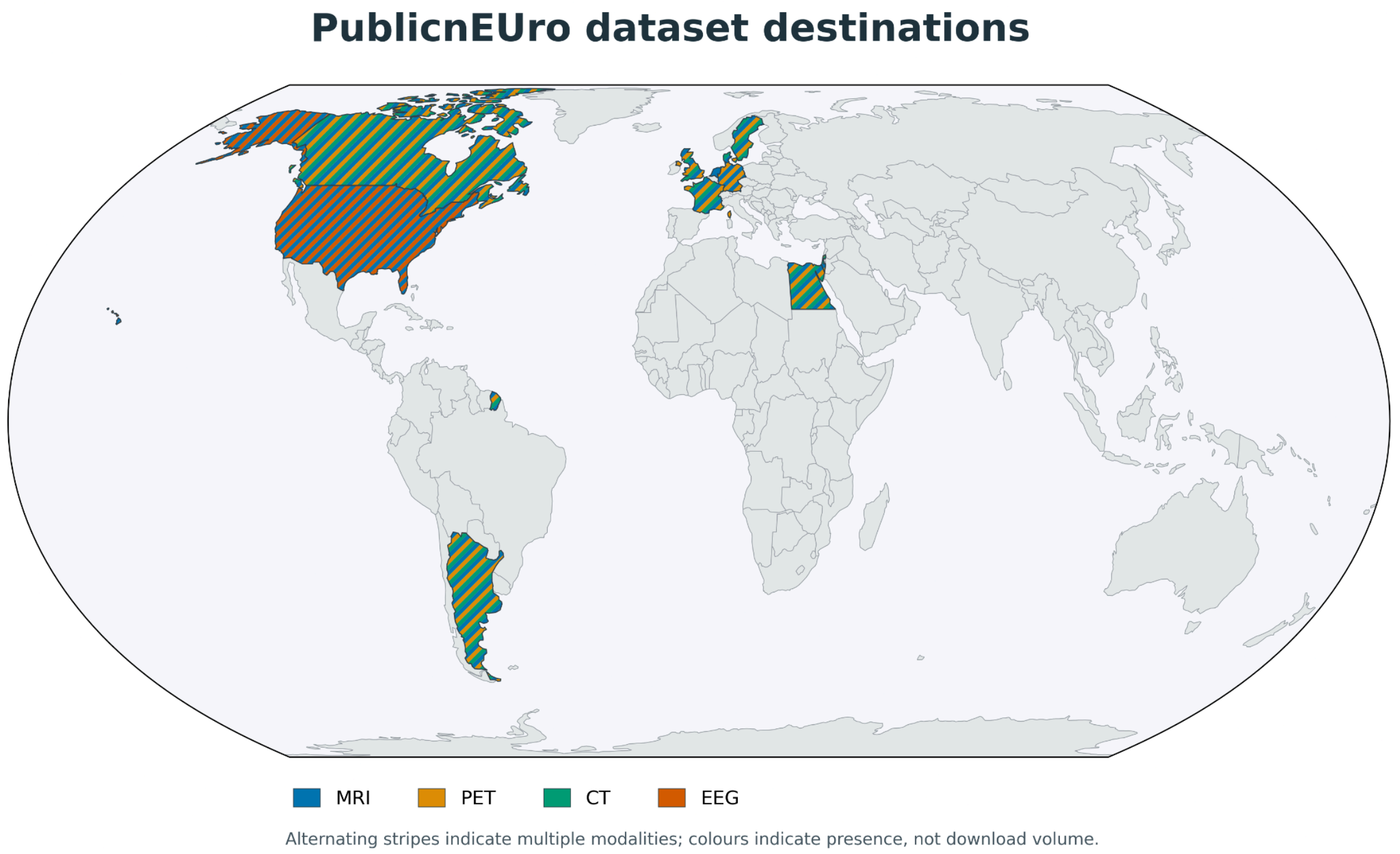

*Figure 2: Geographical distribution of PublicNeuro datasets color coded per imaging modality.*

## 3.1 Assessment

### 3.1.1 FAIR datasets

PublicnEUro offers FAIR sharing (https://doi.org/10.25504/FAIRsharing.4c07be) by adhering to principles beyond 'open access'.

**Findability** is assessed against four principles: (**F1)** (Meta)data are assigned globally unique and persistent identifiers; (**F2)** Data are described with rich metadata; (**F3)** Metadata clearly and explicitly include the identifier of the data they describe; and (**F4)** (Meta)data are registered or indexed in a searchable resource. F1 is achieved for each published dataset with an assigned DOI, including different versions of the same dataset. For F2, F3 and F4, the public catalog and dataset-level JSON records expose rich descriptive, participant, technical, access, lifecycle and citation metadata; each version record explicitly contains its dataset identifier and DOI; and records are discoverable through the searchable catalog, with optional Neurobagel annotation extending federated discovery of participant-level data availability.

**Accessibility** to the data is defined as the ability to retrieve (meta)data by their identifiers using a standardized communication protocol where (**A1.1)** the protocol is open, free, and universally implementable and (**A1.2),** the protocol allows for authentication and authorization procedures, where necessary. In addition, metadata (**A2)** should be accessible even when the data are no longer available. In PublicnEUro, dataset metadata and landing pages are retrievable over HTTPS through DOI resolution and the public catalog. Open data can be downloaded directly, while controlled data use a documented registration, identity verification, dataset-specific DUA and, where required, data-controller approval and international-transfer safeguards. Both browser and command-line retrieval are supported. The preservation policy and metadata model support preservation by maintaining metadata and DOI landing pages independently of file availability across active, archived, retired, withdrawn, and superseded states. This commitment is documented and machine-represented, although the currently published collection does not yet provide a non-active record.

**Interoperability** is defined based on (**I1**) (Meta)data use a formal, accessible, shared, and broadly applicable language for knowledge representation, (**I2)** (Meta)data use vocabularies that follow FAIR principles, and (**I3)** (Meta)data include qualified references to other (meta)data. Here, I1 and I3 were demonstrated, while I2 was only partially demonstrated. PublicnEUro uses formal, machine-readable representations including BIDS, JSON governed by JSON Schema, DUC-compatible JSON, OpenAIRE CERIF XML and re3data XML. Dataset records also make qualified links among the dataset version, DOI, catalog landing page, exact source record, authoritative license or DUA, DUC provenance and, where relevant, replacement versions. Interoperability (I2) is nevertheless incomplete because several DUC condition terms, as well as repository status and retrieval values, are represented as local labels rather than persistent ontology identifiers, and Neurobagel semantic (ontology-based) annotation is encouraged but not mandatory for every dataset.

**Reusable** (Meta)data are richly described data with a plurality of accurate and relevant attributes: (**R1.1)** (Meta)data are released with a clear and accessible usage license, (**R1.2)** (Meta)data are associated with detailed provenance, and (**R1.3)** (Meta)data meet domain-relevant community standards. PublicnEUro records provide descriptive, technical, participant, access, lifecycle (each generated version records its catalog source and hash), and citation attributes; open datasets carry licenses, while controlled datasets retain an authoritative DUA accompanied by a machine-readable DUC summary recording versions, source agreement, and condition-level evidence; and deposited data are organized and validated using the community BIDS standard.

## 3.2 TRUSTed repository

**Transparency**. PublicnEUro's mission, institutional operator, scope, target data types, depositor and user services, open- and controlled-access model, and funding source are available online. Version-controlled policies describe governance, deposit, access, privacy and security, preservation, lifecycle states, sustainability and technical standards. Finally, dataset records expose licenses or DUAs, reviewed DUC, DOI, status and retrieval route. Together, these demonstrate the repository's transparency.

**Responsibility**. Published governance assigns roles to NRU and the Capital Region of Denmark, data controllers, authors, registered users, the Oversight and Management Committees, and the DPO. Data stewardship includes BIDS validation and optional curation, versioned catalog metadata, dataset-version DOIs, controlled access, administration of DUAs and transfer safeguards, access logging, incident response, and persistence of public metadata across lifecycle states. Responsibility for continued custody of participant-level files is explicitly contract- and status-dependent rather than presented as indefinite repository custody.

**User focus.** The service is directed to human brain-imaging depositors and reusers and supports MRI, PET, EEG, MEG and related BIDS data. Users can discover and evaluate available datasets through the public catalog and open metadata, inspect access and reuse conditions before applying, and retrieve eligible data by browser or command line. Depositors receive DOI assignment, controlled-access administration, BIDS validation and optional help with data curation; Neurobagel annotation supports federated discovery. There is no formal public procedure for post-publication data-quality feedback or reuse outcomes currently in place.

**Sustainability**. PublicnEUro has an institutional home at NRU, dataset-level service agreements, a cost-recovery model covering installation, management, maintenance, and storage fees, and explicit transitions between online, cold-archive, external-custody, and unavailable states. The policies cover geo-redundant backup, risk management, contract renewal, return or deletion of files and metadata, DOI persistence, and a repository-closure process.

**Technology**. PublicnEUro combines BIDS and the BIDS Validator, DataLad catalog versioning, Crossref DOIs, JSON Schema, reviewed DUC profiles, and re3data XML exports, and optional Neurobagel annotation. Controlled access uses individual accounts, manual identity verification, password and one-time-password authentication, authorization, and logging. The documented hosting arrangement places files with Computerome in Denmark and specifies encrypted transfer, geo-redundant backups, access controls, risk assessment, and incident response. ISO 27001 applies to identified service components.

## 3.3 OFCT assessment

| Q | Assessment question | PublicnEUro answer and evidence | P |
|---|---|---|---|
| 1 | Does the repository provide access to the data with minimal or no restrictions? | When legally possible, yes. Most datasets are, however, access-controlled to protect sensitive human data. | O |
| 2 | Are users free to download and reuse the data with no or minimal restrictions? | Each dataset has its own terms of use. Some datasets use, for instance, a CC BY license; others have more restrictions on usage in their DUA. | O, A |
| 3 | Does the repository provide a clear license for reuse of the data? | Dataset level: each dataset has either an open license or a dataset-specific DUA. | A, R |
| 4 | Are the data covered by a commons-compliant license? | No, each dataset has its own use terms | O |
| 5 | Does the repository platform make it easy to work with the data? | Yes: eligible data can be retrieved through browser and command-line workflows. | O |
| 6 | Does the repository require or provide documentation to support proper reuse? | Yes: BIDS metadata and documentation are required; licenses or DUAs and reviewed DUC profiles state reuse conditions. | R |
| 7 | Does the repository provide a search facility for data and metadata? | Yes: the public DataLad catalog supports dataset discovery and search. Neurobagel queries provide additional support for participant-level retrieval. | F |

| Q | Assessment question | PublicnEUro answer and evidence | P |
|---|---|---|---|
| 8 | Does the repository assign globally unique and persistent identifiers? | Yes: a Crossref DOI is assigned to every published dataset version. | F |
| 9 | Can users self-assign an identifier to a dataset or contribution? | Not applicable: Q8 answers being Yes. | F |
| 10 | Does the PID or dataset identifier resolve to a landing page describing the data? | Yes: each DOI resolves to a version-specific public catalog record. | F, A, C |
| 11 | Do the metadata explicitly include the identifier of the data they describe? | Yes: dataset-version records include both the PublicnEUro identifier and DOI. | F |
| 12 | Can authors associate an ORCID iD with a dataset? | Supported but not required: the author metadata schema accepts ORCID identifiers. | C |
| 13 | Does the repository support rich metadata for search and reuse? | Yes: records include descriptive, participant, technical, access, lifecycle, provenance and citation metadata. | F |
| 14 | Are the data and metadata associated with detailed provenance? | Yes in full: source records and hashes, metadata sources, version history and DUC mapping evidence are recorded. | R |
| 15 | Does the repository provide the metadata required for data citation? | Yes in full: title, creators, version, DOI, dates and access or license information are provided. | C |
| 16 | Do the metadata include qualified references to other data or metadata? | Yes, in full: records link versions, DOIs, source records, agreements, DUC evidence, and replacements. | I |
| 17 | For data associated with a paper, is bidirectional linkage supported? | Partially: publication identifiers can be represented in machine-readable metadata, but systematic reverse linkage from papers is not assured. | I |
| 18 | Does the repository enforce or allow community standards for data formats or metadata? | Yes: deposited neuroimaging data are organized and validated using BIDS. | R |
| 19 | Does the repository accept metadata applicable to its target community disciplines? | Yes: dataset- and participant-level metadata are supported for human neuroimaging; the dkNET-specific wording was adapted to the PublicnEUro community. | R |
| 20 | Does policy ensure that metadata persist when data are no longer available? | Per stated policy, metadata and DOI landing pages persist across archived, retired, withdrawn, and superseded states. | A |
| 21 | Do the metadata use vocabularies that follow FAIR principles? | At minimum, group-level information contains SNOMED and Disease Ontology IDs. Participant-level information may be tagged using Neurobagel, | I |
| 22 | Does the repository provide a machine-readable landing page? | Not reached because Q10 is Yes. The repository itself has equivalent machine-readable JSON records and XML exports. | A, C |
| 23 | Does the landing page support data citation? | Yes: version-specific DOI and citation metadata are displayed, and machine-readable metadata are exported. | C |
| 24 | Does the repository use a recognized community standard for basic metadata? | Yes: BIDS, JSON Schema, Crossref DOI metadata, OpenAIRE CERIF and re3data representations are used. | R |
| 25 | Can data or metadata be accessed through a standards-compliant API? | Yes: Data can be obtained via our API. All metadata is available as open JSON and XML exports over HTTPS/standard git CLI. | A |

| Q | Assessment question | PublicnEUro answer and evidence | P |
|---|---|---|---|
| 26 | Do metadata use a formal, accessible, shared language for knowledge representation? | Yes: JSON governed by JSON Schema and schema-valid XML representations are public. | I |
| 27 | Does the repository provide API-based search and access to data and metadata? | No search API; static machine-readable records and command-line data retrieval are available; Neurobagel API can be used to query datasets. | F, A |
| 28 | Is repository governance transparent? | Yes: governance, roles, policies, lifecycle commitments, and institutional responsibilities are public and version controlled. | T |
| 29 | Is the repository stakeholder governed? | Partially: Oversight and management structures are documented, but user and depositor participation is not formally incorporated into management. | T |
| 30 | Is the code running the data infrastructure covered by an open-source license? | Partial: user and data database, data catalog, and metadata components use open-source code, but the complete authentication and institutional service stack is not public. | T |
| 31 | Has the repository received a Data Seal of Approval, CoreTrustSeal, or an equivalent certification? | No. PublicnEUro is independently registered and described in re3data (r3d100014932), FAIRsharing (10.25504/FAIRsharing.4c07be) and RRID (SCR_027770) but has not received CoreTrustSeal or equivalent certification. | T |

Table 1. PublicnEUro assessment using the 31-question OFCT instrument. P denotes principle: O, Open; F, Findable; A, Accessible; I, Interoperable; R, Reusable; C, Citable; T, Trustworthy.

# 4. Discussion

PublicnEUro provides a practical model for FAIR discovery and controlled reuse of sensitive neuroimaging data within a European legal and institutional context, but can be leveraged to other issues related to data protection or intellectual property. While it is designed to help EU researchers, datasets from any country can be hosted on PublicnEUro. The FAIR assessment illustrates how public metadata can coexist with protected participant-level data. FAIR principle A1.2 explicitly allows authentication and authorization where necessary, while A2 calls for metadata to remain accessible even when the data are not (Wilkinson et al., 2016). Another prominent neuroimaging data sharing infrastructure, OpenNeuro (Markiewicz et al., 2021) requires data release under a CC0 license and hence provides direct download of its published datasets without requiring an account. PublicnEUro uses *public* in a different sense: anyone can discover and inspect dataset records, identifiers, availability, and access conditions, while retrieval of restricted data requires authorization. However, PublicnEUro also supports openly accessible datasets.

The three assessments performed examine different aspects of PublicnEUro. FAIR evaluates the properties of published data and metadata: whether dataset versions can be found, their access conditions understood, and their contents interpreted and reused. TRUST evaluates the repository responsible for those objects, including its governance, stewardship, responsiveness to users, sustainability, and technology. The 31-question OFCT instrument adds an operational view of observable repository practices, particularly openness and citability. It does not replace the TRUST assessment, whose five principles were not explicitly incorporated into the original instrument. Applying all three also exposes a limitation of the OFCT comparison. Murphy et al. developed and tested their instrument on repositories whose data could be accessed without an approval process; its questions did not address sensitive human-subject data requiring controlled access. PublicnEUro therefore performs differently on OFCT's *Open* questions even when public metadata, persistent identifiers, documented use conditions, and authorized retrieval support FAIR reuse. The results should be interpreted by dimension: unrestricted download, FAIR accessibility, and reliable repository stewardship are distinct properties, and a restriction justified by the data should not be treated as evidence that the other two have failed.

Data sharing depends on infrastructure that must be maintained after the research grant that produced a dataset has ended. Storage is only one expense; repositories also need sustained curation, preservation, security, metadata maintenance and user support (OECD, 2017; Palaiologk et al., 2012). The fMRI Data Center illustrates the risk related to long-term sustainability: changes in funding contributed to the end of full-scale curation and acceptance of new studies, although its existing archive continued in a reduced form (Van Horn & Gazzaniga, 2013). While PublicnEUro's initial development was supported by the Novo Nordisk Foundation, continued operation uses dataset-level service agreements to recover costs from data providers, while discovery and access requests remain free for users. Depending on the dataset, agreements may cover installation, publication, curation, recurring management and storage, supporting the services needed to keep data discoverable and accessible over time. This raises a wider question for data sharing: when researchers seek a repository only after a project has ended, who pays for its continuing stewardship? Free hosting is free to the depositor only because a funder, institution or another contributor covers those costs.

PublicnEUro is now focusing on strengthening its internal procedures by working toward CoreTrustSeal certification (Dillo & Leeuw, 2018) under the 2026–2028 requirements and improving data findability, specifically working toward integration with EBRAINS knowledge graph (Amunts et al., 2022) and by contributing to metadata knowledge exchange.

# Acknowledgement

The Novo Nordisk Foundation (NNF20OC0063277) supported C. Pernet, M. Ganz and the creation of the PublicNeuro repository.